\documentclass[10pt]{IEEEtran}
\usepackage{setspace,amsmath,latexsym,cite,amssymb,epsfig,amsfonts}
\usepackage{graphicx}
\usepackage{psfrag}
\usepackage{footmisc}
\usepackage{multirow}
\usepackage{color}
\usepackage{soul}
\usepackage{multicol}
\usepackage{mathtools}

\twocolumn

\newcommand{\rmd}{\,{\rm d}}

\newtheorem{thm}{Theorem}

\newtheorem{prop}[thm]{Proposition}
\newtheorem{lem}[thm]{Lemma}

        \makeatletter
        \def\fps@eqnfloat{!t}
        \def\ftype@eqnfloat{4}
        
        \newenvironment{eqnfloat*}
               {\@dblfloat{eqnfloat}}
               {\end@dblfloat}
        \makeatother
\title{Secrecy Outage Analysis for Downlink Transmissions in the Presence of Randomly Located Eavesdroppers}
\author{ Gaojie Chen, \IEEEmembership{Member, IEEE}, Justin P. Coon \IEEEmembership{Senior Member, IEEE}, and Marco Di Renzo, \IEEEmembership{Senior Member, IEEE}
\thanks{(c) 2017 IEEE. Personal use of this material is permitted. Permission from IEEE must be obtained for all other users, including reprinting republishing this material for advertising or promotional purposes, creating new collective works for resale or redistribution to servers or lists, or reuse of any copyrighted components of this work in other works. }
\thanks{This work was supported by EPSRC grant number EP/N002350/1 (``Spatially Embedded Networks'').}
\thanks{G. Chen  and J. P. Coon are with the Department of Engineering  Science, University of Oxford, Parks Road, Oxford, UK, OX1 3PJ, Emails: {\tt $\{$gaojie.chen and justin.coon$\}$@eng.ox.ac.uk.}}
\thanks{M. D. Renzo is with the Laboratory of Signals and Systems (L2S), University Paris-Sud XI, France, Emails: {\tt marco.direnzo@lss.supelec.fr.}}
\vspace{1.2mm}\\
\fontsize{10}{10}\selectfont\rmfamily\itshape
}

\begin{document}

\begin{singlespace}
\maketitle
\end{singlespace}

\thispagestyle{empty}
\begin{abstract}
We analyze the secrecy outage probability in the downlink for wireless networks with spatially (Poisson) distributed eavesdroppers (EDs) under the assumption that the base station employs transmit antenna selection (TAS) to enhance secrecy performance.  We compare the cases where the receiving user equipment (UE) operates in half-duplex (HD) mode and full-duplex (FD) mode.  In the latter case, the UE simultaneously receives the intended downlink message and transmits a jamming signal to strengthen secrecy.  We investigate two models of (semi)passive eavesdropping: (1) EDs act independently and (2) EDs collude to intercept the transmitted message.  For both of these models, we obtain expressions for the secrecy outage probability in the downlink for HD and FD UE operation.  The expressions for HD systems have very accurate approximate or exact forms in terms of elementary and/or special functions for all path loss exponents.  Those related to the FD systems have exact integral forms for general path loss exponents, while exact closed forms are given for specific exponents. A closed-form approximation is also derived for the FD case with colluding EDs.  The resulting analysis shows that the reduction in the secrecy outage probability is logarithmic in the number of antennas used for TAS and identifies conditions under which HD operation should be used instead of FD jamming at the UE. These performance trends and exact relations between system parameters can be used to develop adaptive power allocation and duplex operation methods in practice. Examples of such techniques are alluded to herein.

\end{abstract}

\begin{IEEEkeywords}
Physical layer security, stochastic geometry, secrecy outage probability, antenna selection, full-duplex
\end{IEEEkeywords}

\section{Introduction}

Physical layer security, based on Shannon theory using channel coding to achieve secure transmission, has been frequently considered in academia since Wyner's seminal work \cite{A.D75}. Due to the broadcast nature of wireless communications, both the intended receiver and eavesdroppers (EDs) may receive data from the source. But if the capacity of the intended data transmission channel is higher than that of the eavesdropping channel, the data can be transmitted at a rate close to the intended channel capacity so that only the intended receiver can successfully decode the data. This is the principle of physical layer security, where the level of security is quantified by the \emph{secrecy capacity}, i.e., the difference in channel capacities corresponding to the intended data transmission and EDs.

Recently, many works have considered information theoretic security (ITS) over wireless channels, including cooperative relay and jammer networks \cite{E.T08,P.P09}, buffer-added relay networks\cite{G.C14}, multiple-input multiple-output communications (MIMO) \cite{T.L09,H.W09}, full-duplex networks \cite{G.C15}, cognitive radio networks \cite{G.C16}, and distributed beamforming methods \cite{L.D09}. However, all of these works not only assumed a small number of nodes, but also assumed the locations of EDs are known. It is impossible to obtain the location of EDs in practice. For this reason, in 2008, Haenggi provided a powerful method to model the random location distribution of nodes in wireless networks \cite{M.H08,P.C08}.

The impact of random ED locations on secrecy performance has been investigated \cite{X.Z11,G.G14,T.X14,M.G11,T.X15}. The location distribution of EDs can be modeled as a Poisson point process (PPP) or a binomial point process (BPP). In \cite{X.Z11}, the locations of multiple legitimate pairs and EDs were represented as independent two-dimensional PPPs, and the average secrecy throughput in such a wireless network was studied. The MIMO transmission with beamforming was considered later in \cite{G.G14,T.X14} to enhance secrecy performance.

Cooperation is of paramount importance to enhance the capacity and reduce the outage of communication systems subjected to fading and unknown topologies \cite{G.C13}.  As a result, cooperation schemes have been widely applied to enhance communication between legitimate users in a physical layer secrecy context \cite{E.T08,P.P09}. However, relatively little attention has been given to the impact of colluding or cooperative EDs in random spatial networks. Notably, \cite{P.C122} investigated achievable secrecy rates by using the so-called \emph{intrinsically secure graph} formalism, taking into account the effects of ED collusion. Additionally, based on a beamforming technique, the MIMO secrecy connectivity between devices operating in the presence of Rayleigh fading and colluding EDs was analysed in \cite{X.Z111}. However, in that work, the complexity of the system is high due to the use of multiple antennas with beamforming, which may render the system unsuitable for some practical applications.

\begin{table}[!t]\small
\centering
\caption{Notation and Symbols Used in the Paper}
\begin{tabular}{|c|l|}
\hline
   Symbol             &Definition/Explanation         \\ \hline
$\mathbb R^2$         &   two-dimensional space       \\
$\rho_E$              &   density for $\Phi$          \\
$\alpha$              &   path loss exponent          \\
$\epsilon$            &   target secrecy rate         \\
$\mathbb E[\cdot]$    &   expectation operation       \\
$\underset{k\in \{1...K\}}{\max}\left(x_k\right)$              &   maximum function with a set        \\
$[x]^+$               &   $\max(0,x)$                 \\
$\mathbb{P}(\cdot)$   &   probability operator        \\
$G_{s,t}^{m,n}\left(z\left|
\begin{array}{c}
  u_1,\ldots,u_s \\
  v_1,\ldots,v_t
\end{array}
\right.
\right)$              &   Meijer $G$ function         \\
$C^{k}_{K}$           &   binomial coefficient        \\
$\mathbb{Z}^+$        &   positive real numbers       \\
$\Gamma(x)$           &   standard gamma function     \\
$\Gamma(x,y)$         &   upper incomplete gamma function  \\
$\mathcal K_1(x)$     &   first order modified Bessel functions   \\
$O\left(x\right)$     &   big O notation              \\
$F(a,b;c;z)$          &   Gaussian hypergeometric function          \\
$\mathsf{E}_1(x)$     &   exponential integral function          \\
R.V.                &   random variable             \\ \hline
\end{tabular}
\end{table}

\begin{figure}[t]
  \centering
  \centerline{\includegraphics[scale=0.63]{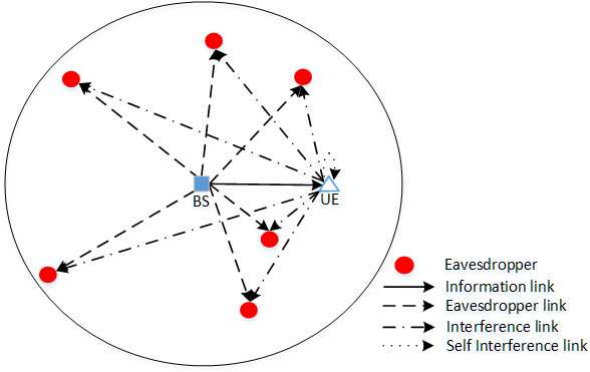}}
 \caption{The wireless network model with randomly located EDs and fixed BS and UE.} \label{fig:symodl}
\end{figure}

In this paper, we analyze the secrecy outage probability in the downlink for wireless networks with randomly (Poisson) distributed EDs.  In order to keep the complexity relatively low at the base station (BS), we consider transmit antenna selection (TAS) rather than beamforming.  Furthermore, we compare the cases where the receiving user equipment (UE) operates in half-duplex (HD) mode and full-duplex (FD) mode.  In the latter case, the UE simultaneously receives the intended downlink message and transmits a jamming signal to disrupt eavesdropping devices \cite{G.C15}.  We also treat the case when EDs act independently as well as the scenario when they collude.  The analytical framework that we present in this paper allows us to make a fair comparison of these four system models (HD/FD and independent/colluding EDs) and thus to draw conclusions about the relative merits and drawbacks of using the secrecy enhancement techniques of TAS and FD jamming under given system parameterizations.  The contributions of the paper are summarized as follows.
\begin{itemize}
  \item We propose TAS at the BS and FD jamming at the receiver to enhance secrecy performance in the presence of randomly located EDs.
  \item We obtain expressions for the secrecy outage probability in the downlink for HD and FD receivers operating in the presence of independent and colluding EDs.  The expressions for HD systems have very accurate approximate or exact forms in terms of elementary and/or special functions for all path loss exponents.  Those related to the FD systems have exact integral forms; exact closed forms are given for certain path loss exponents and closed-form approximations are also derived.
\end{itemize}

The remainder of the paper is organized as follows.  Section \ref{se:II} presents the system model and problem formulation. Sections \ref{se:III} and \ref{se:IV} given an analysis of the secrecy outage probability for the cases where EDs act independently and when they collude, respectively. Section \ref{sec:sim} gives numerical simulations in order to verify the analysis.  Finally, section \ref{sec:con} concludes the paper. \noindent The notation and symbols used in the paper are listed in Table I.

\section{System Model and Secrecy Outage Definition} \label{se:II}

\subsection{System Model}
We consider a secure transmission from the BS to one legitimate UE\footnote{If there are several users in the target cell, only one user is targeted through user scheduling (e.g. random user selection).}. The BS is equipped with $K$ antennas, which it uses to perform TAS in order to maximize the instantaneous signal-to-noise ratio (SNR) at the UE.  The UE is equipped with a hyper-duplex antenna, which can easily switch between HD and FD modes.  Without loss of generality, we locate the BS at the origin in $\mathbb R^2$ and locate the UE at a fixed point a distance $d_{BU}$ along the positive $x$-axis (see Fig.~\ref{fig:symodl}).

We assume EDs are randomly dispersed in a region in the neighbourhood of the BS and the UE.  To this end, we model the EDs as a PPP $\Phi$, which has intensity $\rho_E$ in the closed disk of radius $R$, which we denote by $\mathcal V$, centred at the origin and zero intensity in $\mathbb R^2 \setminus \mathcal V$ (Fig.~\ref{fig:symodl}).  Each ED is equipped with a single antenna, but we consider both the scenarios in which EDs attempt to intercept the downlink signal independently as well as the case when EDs collude to decode the transmitted message.

All channels are assumed to undergo path loss and independent Rayleigh fading effects.  Hence, the coefficient modeling the channel between nodes $i$ and $j$ can be decomposed as $g_{ij} = h_{ij}d^{-\alpha/2}_{ij}$, where $\alpha$ and $d_{ij}$ denote the path loss exponent and the distance between the two nodes, respectively\footnote{In what follows, we set the subscripts $i$ and $j$ to be elements in the set $\{B,U,E\}$ in order to denote transmissions from the BS, UE and EDs, respectively.  For example, $g_{UE_1}$ denotes the channel coefficient between the UE and the first ED in $\Phi$.}. The fading coefficient $h_{ij}$ is modeled as a complex Gaussian random variable with unit variance (i.e., Rayleigh fading is assumed). Therefore, the corresponding channel gains $|g_{ij}|^{2}$ are independently exponentially distributed with mean value $\lambda_{ij}$, and the average channel power is given by $\lambda_{ij}=\mathbb E[|g_{ij}|^{2}] = d_{ij}^{-\alpha}$, where $\mathbb E[\cdot]$ denotes the expectation operation. We assume that the channels are quasi-static, so that the channel coefficients remain unchanged during several packet transmissions but independently vary from coherence time interval to another.

\subsection{Secrecy Performance}
We define downlink secrecy performance using classical wireless wiretap theory. We assume the channel state information (CSI) between the BS and the UE is known by the BS\footnote{This can be achieved by feeding back CSI from the UE to the BS directly or through channel reciprocity in the case of time-division duplex transmissions.}. Therefore, by employing the TAS principle, the BS is able to send a zero-mean symbol $x_s$ with $\mathbb E[|x_s|^2]=1$ to the UE by selecting the $k$th antenna (corresponding to the maximum instantaneous downlink SNR) in a given time slot.

In general, the received signal at the UE can be written as
\begin{equation}\label{1}
   y_{B_kU} = \sqrt{P_B} g_{B_kU} x_s+ \varpi \sqrt{P_U}g_{UU}x_j + n_{U}
\end{equation}
where $P_B$ is the average transmit power at the BS and $n_U$ denotes zero-mean complex Gaussian noise with variance $\sigma_n^2$.  The coefficient $g_{UU}$ corresponds to the residual self-interference channel for the case where FD jamming is employed and $x_j$ denotes the zero-mean jamming signal which has power $\mathbb E[|x_j|^2]=1$.    The average transmit power of the FD UE is $P_U$.  Eq.~(\ref{1}) can be applied to model systems with both HD and FD UEs by adjusting the parameter $\varpi$.  In the HD case, $\varpi = 0$, whereas in the FD case, $\varpi = 1$.

At the same time that the UE receives the message from the BS, the EDs in the set $\Phi$ receive a copy of the transmitted signal.  The received signal at ED $E_e$ can be written as
\begin{equation}\label{2}
   y_{B_kE_e} = \sqrt{P_B} g_{B_kE_e} x_s + \varpi \sqrt{P_U}g_{UE_e}x_j +n_{E_e}
\end{equation}
where $n_{E_e}$ is the Gaussian noise (with variance $\sigma_n^2$) at the ED.

We are interested in quantifying the \emph{secrecy outage probability} in the downlink.  To this end, we require expressions for the BS-UE and BS-ED channel capacities.  Based on the models described above, the capacity of the BS-UE channel can be written as
\begin{equation}\label{3a}
  C_{BU} = \log_2(1+\gamma_{BU})
\end{equation}
where
\begin{equation}\label{5}
\gamma_{BU}=\frac{P_B\underset{k\in \{1...K\}}{\max}\left(\frac{|h_{B_kU}|^2}{d^\alpha_{BU}}\right)}{\varpi P_U|g_{UU}|^2+\sigma_n^2}
\end{equation}
and the $\max$ operation results from the TAS scheme at the BS.  For the BS-ED channel, the capacity is given by
\begin{equation}\label{3b}
C_{BE_*} = \log_2(1+\gamma_{BE_*})
\end{equation}
where
\begin{equation}\label{6}
\gamma_{BE_*}= \mathcal{F}\!\left(\frac{\frac{P_B|h_{B_*E_e}|^2}{d^\alpha_{BE_e}}}{\varpi \frac{P_U|h_{UE_e}|^2}{d^\alpha_{UE_e}}+\sigma_n^2}\right)
\end{equation}
with
\begin{equation}
  B_* = \arg \underset{k\in \{1...K\}}{\max}\left(\frac{|h_{B_kU}|^2}{d^\alpha_{BU}}\right)
\end{equation}
and $\mathcal{F}(\cdot)$ is an operator that takes different forms depending on whether EDs act independently or whether they collude.  In the former case, we have
\begin{equation}\label{eq:F_ind}
  \mathcal F(\cdot) = \max_{e\in\Phi}(\cdot)
\end{equation}
so that we ensure we consider the strongest ED channel, whereas in the case of colluding eavesdroppers, the operator is given by
\begin{equation}\label{eq:F_coll}
  \mathcal F(\cdot) = \sum_{e\in\Phi}(\cdot)
\end{equation}
since all EDs are capable of combining their signals in an optimal manner to decode the message.  Based on these formulae, the secrecy outage probability can be defined as~\cite{P.C121}
\begin{equation} \label{4}
P_{so} = \mathbb{P}([C_{BU} - C_{BE_*}]^+< \epsilon) \simeq \mathbb{P}\left(\frac{\gamma_{BU}}{\gamma_{BE_*}}<\beta\right)
\end{equation}
where $[x]^+=\max(0,x)$, $\mathbb{P}(\cdot)$ denotes the probability operator, $\epsilon$ denotes the target secrecy rate, $\beta = 2^\epsilon$ denotes the target secrecy SNR ratio\footnote{The approximation in (\ref{4}) is a standard assumption for systems operating in the high SNR region.  In this paper, this condition implies $P_B$ is sufficiently large and/or $R$ is sufficiently small.}.

\section{Secrecy Outage Probability for Independently Acting Eavesdroppers} \label{se:III}
Here, we analyse the secrecy outage probability of the downlink for HD and FD UEs under the assumption that EDs act independently of one another.  The EDs cannot share their received signals in this case, so secrecy outage is dictated by the ED with highest channel capacity.  Hence, $\mathcal F(\cdot)$ is defined by (\ref{eq:F_ind}).  We begin by considering an HD UE, then proceed with a treatment of the problem for an FD UE.

\subsection{Half Duplex UE}
Beginning with the right-hand side of \eqref{4}, the secrecy outage probability can be evaluated to yield the result stated in the following proposition.

\begin{prop}\label{prop1}
For large $R$, the downlink secrecy outage probability for an HD UE is, to a good approximation, given by
\begin{multline}\label{10}
P^{(H)}_{so} \simeq 1-\sum_{k=1}^K(-1)^{k+1}C^k_K\frac{\sqrt{pq}}{2^{\frac{p+2q-3}{2}}\pi^{\frac{p+2q}{2}-1}} \\
\times{G}_{0,p+2q}^{p+2q,0}\!\left(\frac{a_k^{2q}b^p}{p^p4^qq^{2q}}\left|
\begin{array}{c}
  - \\
  0,\frac{1}{p},...,\frac{p-1}{p},\frac{1}{2q},\frac{2}{2q},...,1
\end{array}
\right.
\right)
\end{multline}
where $G_{s,t}^{m,n}\left(z\left|
\begin{array}{c}
  u_1,\ldots,u_s \\
  v_1,\ldots,v_t
\end{array}
\right.
\right)$ is the Meijer $G$ function, $C^{k}_{K}=K!/((K-k)!k!)$ is the binomial coefficient, $a_k = kd^\alpha_{BU}$, $b = \pi\rho_E\Gamma(1+{2}/{\alpha})\beta^{2/\alpha}$, $p,q\in \mathbb{Z}^+$ so that $\alpha = p/q$ is a positive rational number, and $\Gamma(x) = \int_0^\infty t^{x-1} e^t\,{\rm d} t$ is the standard gamma function.
\end{prop}
\begin{IEEEproof}
  See Appendix I.
\end{IEEEproof}

Eq.~\eqref{10} provides an explicit, relation between the secrecy outage probability and various system parameters.  A number of interesting points can be noted from this expression.  First, this is the most complete analysis of the HD UE case reported in the literature in that any rational path loss exponent is accounted for in this expression.  Indeed, since the path loss exponent is an experimentally estimated parameter, it is, by definition, rational in practice due to finite precision measurement equipment.  Although the outage probability is given in terms of the Meijer $G$ function, it can be easily evaluated using numerical software such as Mathematica or Maple for any given inputs.  It should be noted that for the special case of $\alpha = 2$, (\ref{10}) reduces to the following expression written in terms of first order modified Bessel functions of the second kind:
\begin{equation}
  P^{(H)}_{so} \simeq1- 2\sum_{k=1}^K(-1)^{k+1}C^k_K \sqrt{a_k b}\, \mathcal K_1\!\left(2 \sqrt{a_k b}\right)
\end{equation}
However, for other values of $\alpha$, the expression given in the proposition is the most compact, accessible form. Note that the expression given in Proposition 1 is independent of $R$. This is because the $R$-dependent terms in the secrecy outage probability expression decay exponentially with $R^\alpha$. (See Appendix I for details.)

For fixed $d_{BU}$, $\rho_E$, $\beta$, and $\alpha$, the secrecy outage probability solely depends on the available number of BS antennas $K$.  It is not a function of the transmit power $P_B$.  This is perfectly intuitive since an increase in $P_B$ would yield a proportional increase in both the UE SNR and the ED SNR.  Thus, in order to satisfy a given secrecy requirement, one must increase the number of antennas used in the TAS procedure.  With large-scale antenna systems and massive MIMO making headlines in the research community in recent years, it is prudent to ask how the secrecy outage probability scales with the number of antennas used for selection.  Since the BS-ED channels are not considered in the selection process, it is clear that the secrecy outage probability decreased monotonically with increasing $K$.  But how fast does this occur?  The following lemma provides some insight to this question.

\begin{lem}
  The downlink secrecy outage probability for an HD UE located in the presence of independently acting EDs is lower bounded by
\begin{equation}
  P_{so}^{(H)} > \frac{\pi \rho_E d_{BU}^2 \beta^{2/\alpha}\Gamma(1+2/\alpha)}{e \,(\ln K)^{2/\alpha}}\left(1 + O\!\left(\frac{1}{(\ln K)^{2/\alpha}}\right)\right)
\end{equation}
as $K\to\infty$.
\end{lem}
\begin{IEEEproof}
  See Appendix II.
\end{IEEEproof}

This result implies that, for large numbers of antennas, secrecy performance improves slowly with increasing $K$.  From a system design perspective, this is a very important result. It suggests that even systems with large numbers of antennas (e.g., massive MIMO systems with a TAS-based secrecy enhancement mode) should exploit only a small subset of independent spatial paths to perform selection.  Such an approach would allow the remaining elements to serve other UEs on separate channels.  The total number of transmit chains (i.e., up-conversion and power amplification circuitry) required would be the number of UEs served in a single channel use.  The actual benefit brought by TAS in the context of enhancing secrecy performance is explored further in Section V through numerical simulations.



\subsection{Full Duplex UE}
In the case where FD jamming is employed by the UE, the jamming signal will affect both the EDs and the UE.  Thus, a self-interference cancellation scheme must be applied at the UE.  Here, we assume the self-interference cancellation scheme is not perfect, and thus residual interference will remain.  Also, we are interested in the \emph{worst-case} secrecy performance.  Thus, in this section, we assume the EDs are interference limited (from the UE's jamming signal).  Mathematically, we set $\sigma_n^2 = 0$.  A similar approach was taken in \cite{X.Z13,C.W15,X.Z131}. Now, beginning with the right-hand side of \eqref{4}, the secrecy outage probability can be evaluated to yield the result stated in the following proposition.

\begin{prop}\label{prop2}
  The downlink secrecy outage probability for an FD UE located in the presence of independently acting EDs  is upper bounded by
\begin{multline}\label{141}
  P^{(F)}_{so} \leq 1-e^{-\rho_E\pi R^2} \sum^K_{k=1}(-1)^{k+1}kC^k_K \\
  \int_0^\infty \frac{\frac{P_U}{d^\alpha_{BU}}(1+\lambda_{UU})+kx\lambda_{UU}}{(\frac{P_U}{d^\alpha_{BU}}+kx\lambda_{UU})^2} \\
  \exp\!\left(\rho_E R^2 \Psi\!\left(\frac{x}{\beta};\alpha,\frac{d_{BU}}{R}\right)-\frac{kd^\alpha_{BU}}{P_U}x\right)\,{\rm d}x
\end{multline}
where
\begin{equation}
  \Psi(y;\alpha,\delta) = \int^{2\pi}_0\int^{1}_0 \frac{yz^{\alpha+1}}{yz^\alpha+({z^2+\delta^2-2z\delta\cos\theta})^{\alpha/2}} \rmd z \rmd \theta
\end{equation}
and $\lambda_{UU} = \mathbb E[|g_{UU}|^2]$ is the average gain of the self-interference channel at the FD UE.
\end{prop}
\begin{IEEEproof}
  See Appendix III.
\end{IEEEproof}


The bound stated above can be evaluated for given sets of parameters by using standard numerical integration techniques or software.  Note that the semi-infinite integral is guaranteed to converge since $\Psi(y;\alpha,\delta)$ is finite for $y\in[0,\infty)$.  For the case where $\alpha = 2$, the bound simplifies somewhat since $\Psi(y;\alpha,\delta)$ evaluates to
\begin{multline}
  \Psi(y;2,\delta) = \frac{\pi  y }{(y+1)^3}\Bigg((y+1)(\psi(y,\delta)-\delta ^2)\\
  + \delta ^2 (y-1) \ln \left(\frac{2 \delta ^2 y}{\delta ^2 (y-1)+(y+1) (\psi(y,\delta)+y+1)}\right)\Bigg)
\end{multline}
where
\begin{equation}
  \psi(y,\delta) = \sqrt{\delta ^4+2 \delta ^2 (y-1)+(y+1)^2}.
\end{equation}
%
%

For fixed $d_{BU}$, $\rho_E$, $\lambda_{UU}$, $\beta$, and $\alpha$, the secrecy outage probability depends on the available number of BS antennas $K$, but also on the UE jamming signal power $P_U$.  This provides two degrees of freedom that can be considered at a system level when determining the best configuration for achieving a target secrecy outage probability.  For example, the UE may locally determine that it should reduce $P_U$ to conserve battery power, which implies the BS should increase the number of antennas used for TAS.  Further analysis of the trade-off between these parameters and the effect this has on system performance are presented in Section V.

\section{Secrecy Outage Probability for Colluding EDs} \label{se:IV}

Here, we analyse the secrecy outage probability in the downlink for HD and FD UEs with the assumption that EDs collude with each other. In contrast to independently acting EDs, colluding EDs can share their eavesdropping information; therefore, all the eavesdropping information can be combined in an effort to decode the downlink message. Under the assumption that optimal combining can be achieved by the EDs, $\mathcal{F}(\cdot)$ is defined by \eqref{eq:F_coll}. We first consider an HD UE, then a treatment of the problem for an FD UE will be provided.

\subsection{Half Duplex UE}
By using the right-hand side of \eqref{4} the secrecy outage probability can written exactly as in Proposition \ref{prop3}.

\begin{prop}\label{prop3}
   The downlink secrecy outage probability for an HD UE located in the presence of colluding EDs is given by
\begin{multline}\label{20}
P^{(H)}_{so} =1 - \sum^K_{k=1}C^k_K(-1)^{k+1}\\
\times \exp\!\left(-\pi R^2\rho_E F\!\left(1,\frac{2}{\alpha};1+\frac{2}{\alpha};-\frac{R^\alpha}{k\beta d_{BU}^\alpha}\right)\right)
\end{multline}
where $F(a,b;c;z)$ denotes the Gaussian hypergeometric function.
\end{prop}
\begin{IEEEproof}
See Appendix IV.
\end{IEEEproof}

Eq. \eqref{20} provides an explicit, exact relation between the secrecy outage probability and various system parameters.  For $\alpha = 2$, this expression simplifies readily to
\begin{equation}
  P^{(H)}_{so} =1 - \sum^K_{k=1}C^k_K(-1)^{k+1} \left(1+\frac{R^2}{\beta d_{\text{BU}}^2 k}\right)^{-\pi \rho_E  \beta d_{{BU}}^2 k}.
\end{equation}
For $\alpha = 4$, (\ref{20}) can be expressed as
\begin{multline}
  P^{(H)}_{so} =1 - \sum^K_{k=1}C^k_K(-1)^{k+1} \\
  \times\exp\!\left({-\pi  \rho_E R  d_{{BU}} \sqrt{\beta k}  \tan ^{-1}\left(\frac{R}{d_{BU}\sqrt{\beta k} }\right)}\right).
\end{multline}
Other values of the path loss exponent do admit closed form expressions by eq. (18). To avoid the redundant discussion, we have not mentioned another pathloss exponents here.



\subsection{Full Duplex UE}

When FD jamming is utilized by the UE, we assume self-interference cancellation is employed by the UE and consider the interference limited regime for EDs (i.e., $\sigma_n^2 = 0$ at each ED).  Following from the right-hand side of \eqref{4}, the secrecy outage probability in this scenario can be evaluated to yield the tight bound stated in the following proposition.

\begin{prop}\label{prop4}
  The downlink secrecy outage probability for an FD UE located in the presence of colluding EDs is bounded by
\begin{multline}\label{22}
P^{(F)}_{so} \leq 1+\sum^K_{k=1}C^k_K(-1)^k \\
\times\exp\!\left(-\rho_E\int_0^{R}\int_0^{2\pi}A_k(r,\theta) e^{A_k(r,\theta)}\mathsf E_1\!\left(A_k(r,\theta)\right)r \rmd\theta \rmd r\right)_,
\end{multline}
where $\mathsf{E}_1(x) = \int^\infty_x\frac{e^{-t}}{t}dt$ denotes the exponential integral and
\begin{equation}
  A_k(r,\theta) = \frac{2k\beta}{P_U} d^\alpha_{BU}\left(\frac{r}{\sqrt{r^2+d_{BU}^2-2rd_{BU}\textrm{cos}(\theta)}}\right)^{-\alpha}.
\end{equation}
\end{prop}
\begin{IEEEproof}
See Appendix V.
\end{IEEEproof}
Eq. \eqref{22} can be evaluated for given sets of parameters by using standard numerical integration techniques or software.  However, it is useful to have an approximation of this expression that does not require numerical integration.  We give such an approximation for $\alpha = 2$ in the following lemma, and we validate the approximation in the next section through an extensive simulation study.
\begin{lem}\label{lemma2}
  For $\alpha = 2$, the downlink secrecy outage probability for an FD UE located in the presence of colluding EDs operating in the interference limited regime is approximated by
\begin{multline}\label{222}
P^{(F)}_{so} \simeq 1+\sum^K_{k=1}C^k_K(-1)^k \exp\!\Big(-\rho_E\big(\pi\varrho^2 \\
-\frac{\pi P_U}{2k\beta} \left(({\varrho}/{d_{BU}})^2 -\ln({1-({\varrho}/{d_{BU}})^2})\right)+\Omega(\beta;d_{BU},R,A_0)\big)\Big)
\end{multline}
where $\varrho\in(0, R)$, $A_0 = {2k\beta} d^2_{BU} /{P_U}$, and $\Omega(\beta;d_{BU},R,A_0)$ is given as \eqref{omega} at the top of the next page.
\end{lem}
\begin{figure*}
\begin{equation}\label{omega}
\begin{split}
\Omega(\beta;d_{BU},R,A_0) = & - \frac{A_0\pi}{\varrho^4R^4}
(4 R^4\varrho^4 d_{BU}^2(A_0+1/4)(\ln (\varrho)^2 + 2\ln (A_0 d_{BU}) \ln (R/\varrho) - \ln (R)^2) \\
&+R^4\varrho^4 \ln \varrho\,((A_0+1)\varrho^2  -8(A_0\kappa-(9/4)A_0^2+(1/4)\kappa+1/4)d_{BU}^2 -d_{BU}^4A_0/\varrho^4)  \\
&-R^4 \varrho^4 \ln R\,((A_0+1)R^2 -8(A_0\kappa-(9/4)A_0^2+(1/4)\kappa+1/4)d_{BU}^2 -d_{BU}^4A_0)\\
&+(R^2-\varrho^2)(\varrho^2(R^2(A_0+1)\varrho^2 +d_{BU}^4A_0)R^2\ln (A_0) +\varrho^2(R^2(A_0+1)\varrho^2+d_{BU}^4A_0)R^2\ln (d_{BU})\\
&+R^4(-A_0^2+(\kappa+1)A_0+\kappa+3/2)\varrho^4+A_0((\kappa-9A_0)R^2-(1/2)d_{BU}^2A_0)d_{BU}^4\varrho^2 -(1/2)R^2 d_{BU}^6A_0^2)).
\end{split}
\end{equation}
\hrule
\end{figure*}
\begin{IEEEproof}
See Appendix V.
\end{IEEEproof}




\section{Simulations Results} \label{sec:sim}
In this section, simulation results (based on the left-hand side of \eqref{4}) are given to verify the above analysis. In the simulations, we assume the noise variance $\sigma_{n}^{2} = 1$, the transmission-power-to-noise ratio $P_B/\sigma^2_n = 50$ dB, and the target secrecy SNR $\beta = 1$. The simulation results are obtained by averaging over $10^5$ independent Monte Carlo trials. Moreover, the single-antenna scheme ($K = 1$) is our benchmark and has been considered in this section.

\begin{table}[!t]\small
\centering
\caption{Effects of parameters increases on secrecy outage probability. Upward (downward) arrows signify an increase (decrease). Horizontal dashes denote little to no change. An arrow followed by a dash signifies convergence to a positive, finite value. Arrows followed by parenthetical expressions denote the trend of increase/decrease (either logarithmic or a power law).}
\begin{tabular}{|c|c|c|c|c|}
\hline
                & HD IE& FD IE  & HD CE   & FD CE               \\ \hline
$K$  $\nearrow$    &   $\searrow$ (log) &  $\searrow$ (log)  & $\searrow$ (log) & $\searrow$ (log)  \\ \hline
$\rho_E$ $\nearrow$ &  $\nearrow$ &  $\nearrow$   & $\nearrow$  & $\nearrow$   \\ \hline
$\beta$  $\nearrow$ &  $\nearrow$ &  $\nearrow$   & $\nearrow$  & $\nearrow$   \\ \hline
$d_{BU}$  $\nearrow$ &  $\nearrow$ &  $\nearrow$   & $\nearrow$  & $\nearrow$   \\ \hline
$\alpha$  $\nearrow$&  $\searrow$~~$-$ &  $\nearrow$~~$-$  & $\searrow$~~$-$  & $\nearrow$~~$-$   \\ \hline
$\lambda_{UU}$  $\nearrow$&  $-$ &  $\nearrow$   & $-$  & $\nearrow$   \\ \hline
$P_U/\sigma_n^2$  $\nearrow$   &  $-$ &  $\searrow$ (power)   & $-$  & $\searrow$ (power)   \\ \hline
$P_B/\sigma_n^2$  $\nearrow$   &  $-$ &  $-$   & $-$  & $-$   \\ \hline
\end{tabular}
\end{table}

Firstly, Table I gives an overview of how different system parameters affect secrecy outage for the four cases discussed in the previous sections, where IE and CE denote independent and colluding eavesdropper case, respectively, $\nearrow$, $\searrow$ and $-$ denote increasing, decreasing and unchanging trends, respectively. It is clear that the secrecy outage probability for each of the four cases increases with increasing ED density, target SNR $\beta$, and BS-UE distance $d_{BU}$. On the contrary, the secrecy outage probability decreases with the number of transmission antennas $K$. With the increasing of $\alpha$, the secrecy outage probability in the HD case decreases slowly while the secrecy outage probability in the FD case increases steadily until it converges to a finite value (more details in Fig. \ref{fig:Fig6}). Note that the secrecy outage probability is independent of the transmit power-to-noise ratio $P_B/\sigma^2_n$ for the BS. Finally, the transmit power-to-noise ratio $P_U/\sigma^2_n$ for the UE and the residual self-interference channel gain ($\lambda_{UU}$) only affects the FD case which is shown in Figs. \ref{fig:Fig4} and \ref{fig:Fig5}, respectively.

\begin{figure}[!t]
\centering
\includegraphics[scale=0.6]{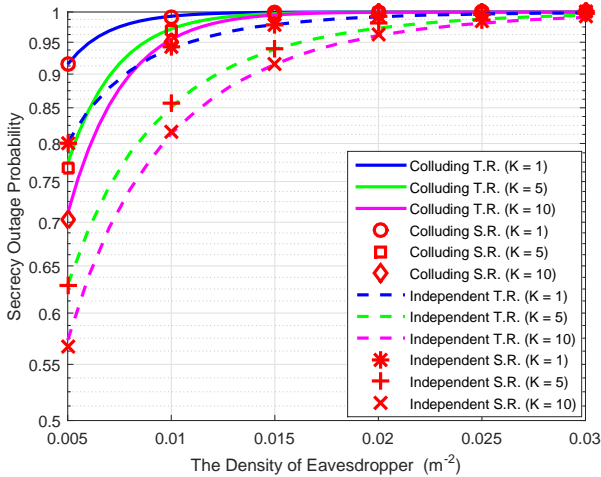}
\caption{\small Theoretical (T.R.) {\em vs} simulated (S.R.) secrecy outage probabilities for the HD UE in the presence of different densities of EDs, where $\alpha = 4$, $d_{BU} = 10$ m and $R = 100$ m.}
\label{fig:Fig2}
\end{figure}

\begin{figure}[!t]
\centering
\includegraphics[scale=0.6]{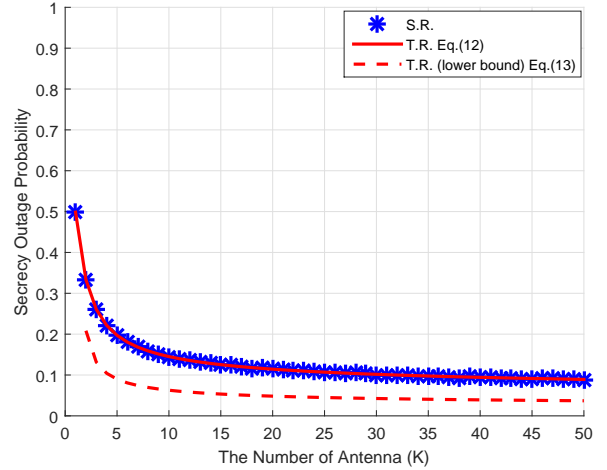}
\caption{\small The comparison of secrecy outage probabilities for HD UEs with different numbers of antennas (K), where $\rho_E = 0.005$ $\textrm{m}^{-2}$, $\alpha=2$ $d_{BU} = 5$ m and $R = 50$ m.}
\label{fig:Fig22}
\end{figure}

Fig. \ref{fig:Fig2} verifies the secrecy outage probabilities for the HD UE for independent EDs \eqref{10} and colluding EDs \eqref{20}, respectively. Here we let $d_{BU} = 10$ m, $R = 100$~m and $\alpha = 4$. Both the simulated results (S.R.) and theoretical results (T.R.) are presented, which are shown to perfectly match. Furthermore, it is clear from these results that the secrecy outage probability slowly decreases as the number of transmit antennas increases for both cases, which has been predicted by Lemma 2. The secrecy outage probability for independent EDs is always smaller than that for the colluding case, because of the shared eavesdropping information.

Fig. \ref{fig:Fig22} compares secrecy outage probabilities for HD UEs with different numbers of antennas (K), where $\rho_E = 0.005$ $\textrm{m}^{-2}$, $\alpha=2$ $d_{BU} = 5$ m and $R = 50$ m. It is clear to see that when the number of antennas ranges from 1 to 15, there exists a significant secrecy performance gain. However, with increasing numbers of antennas after 15, secrecy performance improves slowly with $1/\textrm{ln}(K)$, which has been confirmed by \textit{Lemma} 2. From a system design perspective, this is a very important result. It suggests that even systems with large numbers of antennas (e.g., massive MIMO systems with a TAS-based secrecy enhancement mode) should exploit only a small subset of independent spatial paths to perform selection.

\begin{figure}[t]
\centering
\includegraphics[scale=0.6]{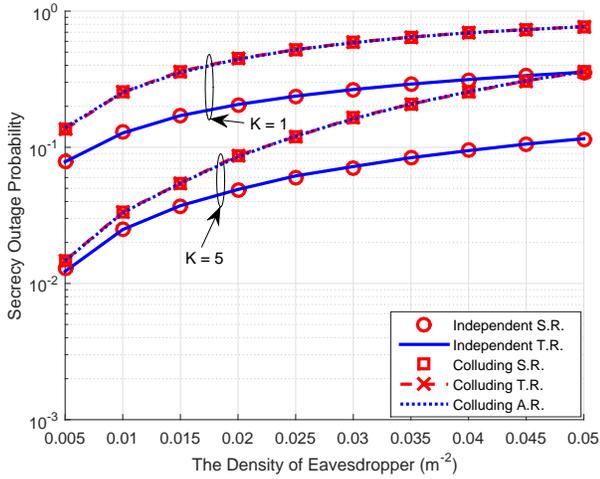}
\caption{\small T.R. {\em vs} S.R. and approximation results (A.R.) secrecy outage probabilities for the FD UE in the presence of different densities of EDs, where $\alpha = 2$, $d_{BU} = 5$ m, $R = 50$ m and $\varrho = 1$ for A.R.}
\label{fig:Fig3}
\end{figure}

\begin{figure}[!htb]
\centering
\includegraphics[scale=0.6]{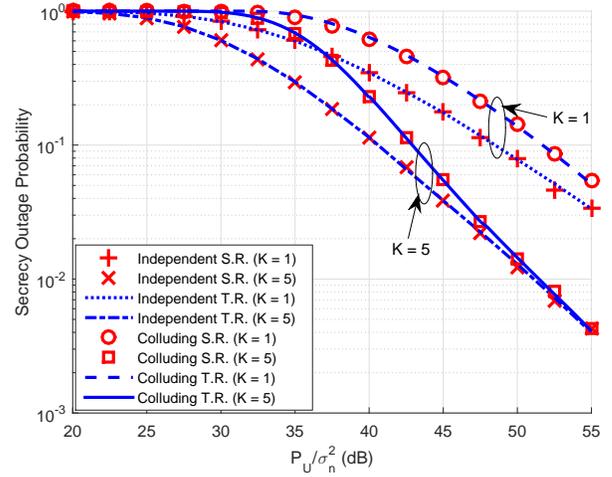}
\caption{\small T.R. {\em vs} S.R. secrecy outage probabilities for the FD UE with different transmission power-to-noise ratios at the UE, where $d_{BU} = 5$ m, $R = 50$ m and $\rho_E = 0.005~\textrm{m}^{-2}$.}
\label{fig:Fig4}
\end{figure}

\begin{figure*}[]
\begin{minipage}[b]{0.5\linewidth}
  \centering
  \centerline{\includegraphics[scale=0.6]{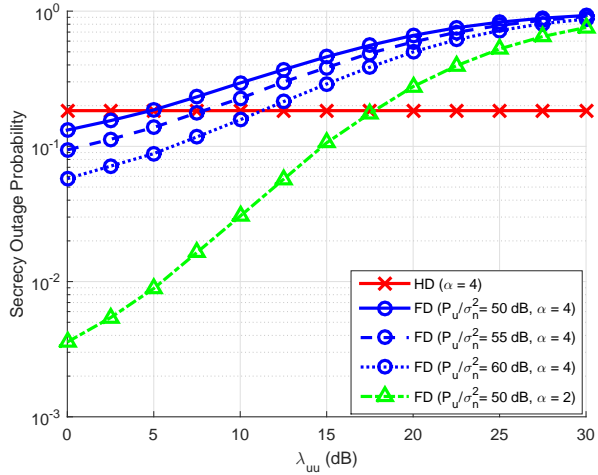}}
 \vspace{0.3cm}
  \centerline{\small (a) Independent EDs}
\end{minipage}
\hfill
\begin{minipage}[b]{0.5\linewidth}
  \centering
  \centerline{\includegraphics[scale=0.6]{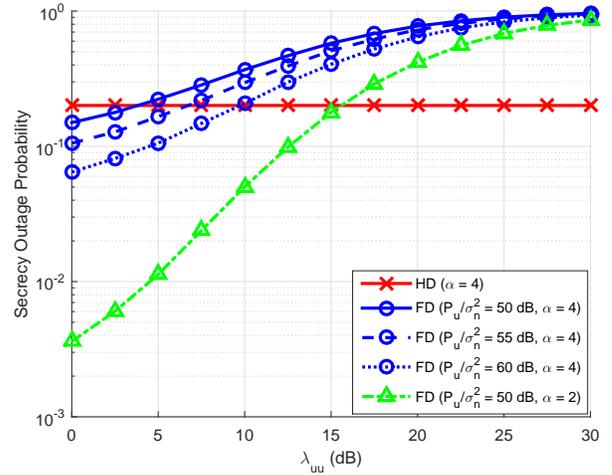}}
  \vspace{0.3cm}
  \centerline{\small (b) Colluding EDs}
\end{minipage}
 \caption{\small The comparison of secrecy outage probabilities for FD and HD UEs with different residual self-interference channel gains, where $d_{BU} = 10$ m, $R = 50$ m and $\rho_E = 0.001~\textrm{m}^{-2}$.} \label{fig:Fig5}
\end{figure*}

\begin{figure}[t]
\centering
\includegraphics[scale=0.6]{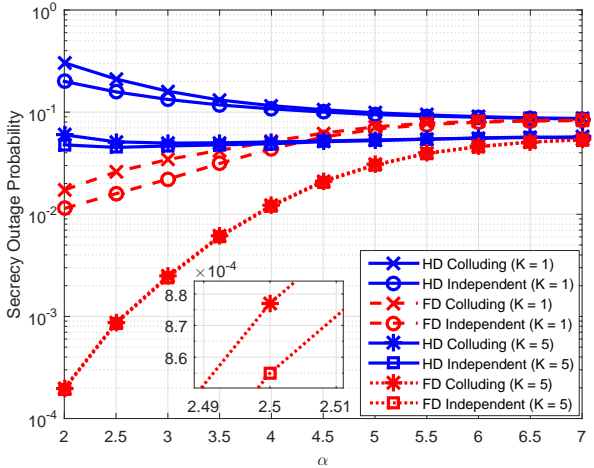}
\caption{\small The comparison of secrecy outage probabilities for FD and HD UEs with different pathloss exponents, where $\rho_E = 0.001$ $\textrm{m}^{-2}$, $\lambda_{UU} = 0$ dB, $d_{BU} = 5$ m and $R = 50$ m.}
\label{fig:Fig6}
\end{figure}

The comparison between the T.R. and S.R. of secrecy outage probabilities for the FD UE is shown in Fig. \ref{fig:Fig3}, where we let $\lambda_{UU} = 0$ dB, $d_{BU} = 5$ m, $R = 50$ m, $\varrho = 1$\footnote{According to the simulation results, accurate results were obtained for $\varrho$ close to one.} and $\alpha = 2$. Again, the theoretical results generated with the help of \eqref{141} for independent EDs and \eqref{22} for colluding EDs are well matched to the simulation results. And the approximation results (A.R.) \eqref{222} for colluding EDs were confirmed by simulation results as well. Moreover, it is clear that the secrecy outage probability decreases exponentially quickly as the density of EDs decreases, as predicted by Propositions 3 and 5.

Fig. \ref{fig:Fig4} shows the comparison between the T.R. and S.R. of secrecy outage probabilities versus different transmission power-to-noise ratio for the FD UE in the presence of independent and colluding EDs, where $\lambda_{UU} = 0$ dB, $d_{BU} = 5$ m, $R = 50$ m, $\rho_E = 0.005~\textrm{m}^{-2}$ and $\alpha = 2$. For these system parameters, the average number of EDs located in the vicinity of the BS (i.e., the circle of radius $R$ centered at the BS) is approximately 39. Hence, these parameters provide a view of performance in a fairly hostile environment. We can see that the T.R. of independent \eqref{141} and colluding \eqref{22} EDs are well matched to the S.R. Then it is clear that the secrecy outage probability linearly decreases asymptotically on the log-log scale as the transmission power-to-noise ratio at the UE increases for both cases. Furthermore, when the required secrecy outage probability is 0.05, if the number of antennas increases from 1 to 5, almost 10 dB SNR can be saved for both cases.  The above figures verified the analysis in Section III and IV. In order to maintain clarity of presentation, only the simulation results are shown in the following figures.


According to \cite{S.H14}, radio transmissions always encounter a bandwidth constraint that limits maximum self-interference cancellation.  Therefore, it is useful to consider how residual self-interference affects the secrecy outage performance of the FD scheme. Fig.~\ref{fig:Fig5} compares the secrecy outage probabilities of independent (Fig.~\ref{fig:Fig5}(a)) and colluding  (Fig.~\ref{fig:Fig5}(b)) EDs for the HD and FD modes with respect to different $\lambda_{UU}$ and $\alpha$, where $d_{BU} = 10$ m, $R= 50$ m, $K = 5$ and $\rho_E = 0.001~\textrm{m}^{-2}$. Hence, in this example, we consider a more secure environment with an average of about eight EDs located in the vicinity of the BS. It is clearly shown in the figures that as the residual self-interference increases, the secrecy outage probability of the FD case is adversely affected. Obviously, there is no self-interference for the HD scheme; hence, the performance is constant for all $\lambda_{UU}$ in this figure. Of more interest is the observation that the secrecy outage probabilities of the HD mode are always less than for the FD mode when $\lambda_{UU}$ is less than about 11.5 dB and 10 dB for independent and colluding cases, respectively, when $P_U/\sigma^2_{U} = 60$ dB.  Furthermore, an important point shown in Fig.~\ref{fig:Fig5} is that when the path loss exponent $\alpha$ increases, the enhancement of secrecy performance by using the FD scheme will be limited due to the significant attenuation of the jamming signal from the FD UE to the EDs. Therefore, we should increase the jamming power $P_U$ according to the theoretical expressions given in Propositions 3 and 5 so that the secrecy outage probability can be reduced. This information can be employed in practice to switch between HD and FD modes given the bandwidth constraints of the system with different path loss exponents. Since the available system bandwidth of modern communication links can change based on channel quality and the prescribed quality of service, this observation could be of great importance in future cellular networks \cite{S.H14}.

Fig. \ref{fig:Fig6} shows the comparison of secrecy outage probabilities versus different path loss exponents for the HD and FD UE cases operating in the presence of independent and colluding EDs, where $\lambda_{UU} = 0$ dB, $d_{BU} = 5$ m, $R = 50$ m, $\rho_E = 0.001~\textrm{m}^{-2}$ and $K = 1~\textrm{and}~5$. In this example, there are on average about eight eavesdroppers in the vicinity of the network. We can see that the secrecy outage probability for HD UE with independent and colluding EDs slightly decreases until reaching a flat tail with an increasing path loss exponent. On the contrary, the secrecy outage probability for the FD case increases to this saturation point. The reason is that when the UE's transmission power fixed, the power of the jamming signal from the FD UE is attenuated significantly for large $\alpha$. Furthermore, it is clear that the secrecy outage probability for colluding EDs is always higher than for independent EDs.

\section{Conclusion}\label{sec:con}
In this paper, we studied a method of enhancing secrecy performance in wireless networks with randomly located independent and colluding EDs, which relies on the use of TAS at the base station and an FD jamming scheme at the UE. For both of these models, we obtained expressions for the secrecy outage probability in the downlink for HD and FD UE operation.  The expressions for HD systems have very accurate approximate or exact forms in terms of elementary and/or special functions for all path loss exponents.  Those related to the FD systems have very accurate approximate or exact integral forms for general path loss exponents, while exact closed forms are given for specific exponents.  These results have been confirmed by simulated simulations which showed how secrecy performance can be enhanced by TAS and FD communications. Our results provide useful insight and analytical tools that can be used to develop adaptive system solutions (examples were briefly discussed for hybrid HD/FD UE operation) as well as a solid basis for further study.

\section*{Appendix I}

We assume all channels are independent and identically distributed (i.i.d.); consequently, the cumulative distribution function (CDF) and probability density function (PDF) of $\gamma_{BU}$ in \eqref{5} with $\varpi = 0$ are given by
\begin{equation}\label{8}
\begin{split}
&F_{\gamma_{BU}}(x) = \left(1-e^{-xd^\alpha_{BU}}\right)^K = \sum^K_{k=0}C^k_K(-1)^ke^{-kxd^\alpha_{BU}},\\
&f_{\gamma_{BU}}(x) = \sum_{k=1}^KC^k_K(-1)^{k+1}kd^\alpha_{BU} e^{-kxd^\alpha_{BU}},
\end{split}
\end{equation}
respectively, where $C^{k}_{K}=K!/[k!(K-k)!]$ is the binomial coefficient. Then, the CDF of $\gamma_{BE_*}$ in \eqref{6} with $\varpi = 0$ can be calculated as
\begin{equation} \label{9}
\begin{split}
&F_{\gamma_{BE_*}}(y) = \mathbb{P}\left(\underset{e\in \Phi}{\max}\left(\frac{|h_{B_*E_e}|^2}{d^\alpha_{BE_e}}\right)<y\right)\\
& \overset{(a)}{=}  E_\Phi\left[\prod_{e\in \Phi}\mathbb{P}\left(|h_{B_*E_e}|^2<yd^{\alpha}_{BE_e}\mid \Phi\right)\right]\\
& =  E_\Phi\left[\prod_{e\in \Phi}\left(1-e^{-yd^{\alpha}_{BE_e}}\right)\right]\\
& \overset{(b)}{=} \textrm{exp}\left(-\rho_{E}\int^{2\pi}_0\int^{R}_0r\left(e^{-yr^{\alpha}}\right)\rmd r\rmd \theta\right)\\
& \overset{(c)}{=} \textrm{exp}\left(-\frac{2\pi\rho_E}{\alpha y^{\frac{2}{\alpha}}}\left(\Gamma\left(\frac{2}{\alpha}\right)-\Gamma\left(\frac{2}{\alpha}, yR^\alpha\right)\right)\right)\\
& \overset{(d)}{=} \textrm{exp}\left(-\frac{2\pi\rho_E}{\alpha y^{\frac{2}{\alpha}}}\Gamma\left(\frac{2}{\alpha}\right)\right)\left(1+\frac{2\pi\rho_E}{\alpha y^{\frac{2}{\alpha}}}O(R^{2-\alpha}y^{2/\alpha-1}e^{-yR^\alpha})\right),
\end{split}
\end{equation}
where $\Gamma(\cdot)$ and $\Gamma(\cdot,\cdot)$ denote the gamma and upper incomplete gamma function, respectively, and where eq. (a) follows from the independence of R.V.s $\{|h_{B_*E_e}|^2; E_e\in\Phi\}$; eq. (b) holds for the probability generating functional lemma \cite{M.H12}; eq. (c) holds by using eq. (3.326.4) in \cite{I.S07}; eq. (d) follows from the asymptotic expansion of the incomplete gamma function ($R\to\infty$) \cite{M.A72}.

According to the definition of secrecy outage probability in \eqref{4}, \eqref{8} and \eqref{9}, we can obtain an approximation of the secrecy outage probability as follows
\begin{equation}
\begin{split}
&P^{(H)}_{so} =1- \int_0^\infty f_{\gamma_{BU}}(x)F_{\gamma_{BE_*}}\left(\frac{x}{\beta}\right)\rmd x\\
& =1-\sum_{k=1}^KC^k_K(-1)^{k+1}kd^\alpha_{BU}\int_0^\infty e^{-kxd^\alpha_{BU}}e^{-\frac{2\pi\rho_E}{\alpha\left(\frac{x}{\beta}\right)^{2/\alpha}}\Gamma(\frac{2}{\alpha})}\rmd x.\\
\end{split}
\end{equation}
We let
\begin{equation}
\begin{split}
I = \int_0^\infty e^{-ax}e^{-\frac{b}{x^c}}\rmd x = \int_0^\infty ue^{-au} e^{-\left(\frac{b^{1/c}}{x}\right)^c} \frac{\rmd u}{u}
\end{split}
\end{equation}
where $a = kd^\alpha_{BU}$, $b = \frac{2\pi\rho_E}{\alpha}\Gamma(\frac{2q}{p})\beta^{2q/p}$ and $c = 2q/p$. By using the Mellin convolution theorem, we can get the Mellin transform as
\begin{equation}
\begin{split}
\mathcal{M}[I;s] = \frac{p}{2qa^{s+1}}\Gamma\left(\frac{ps}{2q}\right)\Gamma(1+s).
\end{split}
\end{equation}
Then the inverse transform can be written as
\begin{equation}
\begin{split}
&I = \frac{p}{2\pi ia}\int_{u-i\infty}^{u+i\infty}\Gamma\left(ps\right)\Gamma\left(2q(s+\frac{1}{2q})\right)(a^{2q}b^{p})^{-s}\rmd s\\
& \overset{(a)}{=}\frac{\sqrt{pq}}{a2^{\frac{p+2q-3}{2}}\pi^{\frac{p+2q}{2}-1}}\frac{1}{2\pi i}\\
&\times\int_{u-i\infty}^{u+i\infty}\left( \frac{a^{2q}b^p}{p^p4^qq^{2q}}\right)^{-s}\prod_{n=0}^{p-1}\Gamma\left(s+\frac{n}{p}\right)\prod_{n=0}^{2q-1}\Gamma\left(s+\frac{1+n}{2q}\right)\rmd s\\
& = \frac{\sqrt{pq}}{a2^{\frac{p+2q-3}{2}}\pi^{\frac{p+2q}{2}-1}} \\&~~~\times{G}_{0,p+2q}^{p+2q,0}\!\left(\frac{a_k^{2q}b^p}{p^p4^qq^{2q}}\left|
\begin{array}{c}
  - \\
  0,\frac{1}{p},...,\frac{p-1}{p},\frac{1}{2q},\frac{2}{2q},...,1
\end{array}
\right.
\right),
\end{split}
\end{equation}
where $G(\cdot)$ denotes Meijer's G furcation, $u>0$ and (a) holds from the multiplication theorem \cite{M.A72}.

\section*{Appendix II}\label{app2}
We begin with the following basic integral definition of the secrecy outage probability for this case
\begin{equation}
  P_{so}^{(H)} = b_1 c_1\int_{0}^{\infty}(1-e^{-a x})^K\frac{e^{-b_1/x^{c_1}}}{x^{{1+c_1}}} \rmd x
\end{equation}
where $a = \beta d_{BU}^\alpha$, $b_1 = c_1\pi \rho_E \Gamma(2/\alpha)$ and $c_1 = 2/\alpha$.  This expression can easily be derived from the definitions of the UE SNR and the ED SNR and follows the calculations presented in Appendix I.  Since the integrand is nonnegative on the interval $[0,\infty)$, we have the simple relations
\begin{align}
  P_{so}^{(H)} &> b_1 c_1\int_{\frac{\ln K}{a}}^{\infty}(1-e^{-a x})^K\frac{e^{-b_1/x^{c_1}}}{x^{{1+c_1}}} \rmd x \nonumber \\
  & > b_1 c_1 \left(1-\frac 1 K\right)^K\int_{\frac{\ln K}{a}}^{\infty}\frac{e^{-b_1/x^{c_1}}}{x^{{1+c_1}}} \rmd x  \nonumber \\
  & = \left(1-\frac 1 K\right)^K\left(1 - \exp\left(-\frac{a^{c_1} b_1}{(\ln K)^{c_1}}\right)\right)
\end{align}
where the equality results from the substitution $u = 1/x^{c_1}$.  Letting $K$ grow large, the final line of the equation given above becomes
\begin{equation}
 e^{-1}\left(1 + O\!\left(\frac 1 K\right)\right)
   \left(1 - \left(1-\frac{a^{c_1} b_1}{(\ln K)^{c_1}}+ O\!\left(\frac{1}{(\ln K)^{2c_1}}\right)\right)\right)
\end{equation}
and the result stated in the lemma follows.

\section*{Appendix III}
According to \eqref{4}, \eqref{5} and \eqref{6} with $\varpi = 1$, we let $X_1 = P_U\underset{k\in (1...K)}{\max}(|h_{B_kU}|^2)$ and $X_2 = |h_{UU}|^2$. Then after self-interference cancellation, the average channel gain of the residual self-interference can be denoted as $\lambda_{UU}$. Therefore, the CDF of $X_1$  and the PDF of $X_2$ can be written as
\begin{equation} \label{12}
\begin{split}
&F_{X_1}(x_1) = \sum^K_{k=0}C^k_K(-1)^ke^{-\frac{kx_1d_{BU}^{\alpha}}{P_U}}\\
&f_{X_2}(x_2) = 1/\lambda_{UU}e^{-x_2/\lambda_{UU}},
\end{split}
\end{equation}
respectively. The CDF and PDF of $X = \frac{X_1}{X_2+1}$ are given by
\begin{equation} \label{13}
\begin{split}
F_{X}(x) &=\int_0^\infty F_{x_1}(x(x_2+1))f_{x_2}(x_2)\rmd x_2\\
& =\sum^K_{k=0}C^k_K(-1)^k\frac{\frac{P_U}{d^\alpha_{BU}}e^{-\frac{kxd^\alpha_{BU}}{P_U}}}{\frac{P_U}{d^\alpha_{BU}}+kx\lambda_{UU}}
\end{split}
\end{equation}
and 
\begin{equation} \label{14}
\begin{split}
f_{X}(x) =&\sum^K_{k=1}C^k_K(-1)^{k+1}\\
&\frac{(P_U+kx\lambda_{UU}d^\alpha_{BU}+P_U\lambda_{UU})ke^{-\frac{kxd^\alpha_{BU}}{P_U}}}{d^\alpha_{BU}(\frac{P_U}{d^\alpha_{BU}}+kx\lambda_{UU})^2}.
\end{split}
\end{equation}

Then letting $Y = \underset{e\in \Phi}{\max}\left(\frac{|h_{B_*E_e}|^2}{d^\alpha_{BE_e}}/\frac{|h_{UE_e}|^2}{d^\alpha_{UE_e}}\right)$, it is possible to show that the CDF of $Y$ can be written as (37) in the top of the next page,
\begin{figure*}
\begin{equation} \label{16}
\begin{split}
F_{Y}(y) &= \mathbb{P}\left(\underset{e\in \Phi}{\max}\left(\frac{\frac{|h_{B_*E_e}|^2}{d^\alpha_{BE_e}}}{\frac{|h_{UE_e}|^2}{d^\alpha_{UE_e}}}\right)<y\right)= E_\Phi\left[\mathbb{P}\left(\underset{e\in \Phi}{\max}\left(\frac{|h_{B_*E_e}|^2d^{-\alpha}_{BE_e}}{|h_{UE_e}|^2d^{-\alpha}_{UE_e}}\right)<y\mid \Phi\right)\right]\\
&\overset{(a)}{=} E_\Phi\left[\prod_{e\in \Phi}\mathbb{P}\left(\frac{|h_{B_*E_e}|^2}{|h_{UE_e}|^2}<y\frac{d^{\alpha}_{BE_e}}{d^{\alpha}_{UE_e}}\mid \Phi\right)\right] \overset{(b)}{=} E_\Phi\left[\prod_{e\in \Phi}\left(\frac{yd_{BE_e}^\alpha}{yd_{BE_e}^\alpha+d_{UE_e}^\alpha}\right)\right]\overset{(c)}{=} \textrm{exp}\left(-\rho_E\int^{R}_0\int^{2\pi}_0r\Xi(y;r,\theta)\rmd\theta \rmd r\right).
\end{split}
\end{equation}
\hrule
\end{figure*}
where
\begin{equation}
  \Xi(y;r,\theta) = 1-\frac{yr^\alpha}{yr^\alpha+(\sqrt{r^2+d^2_{BU}-2rd_{BU}\textrm{cos}\theta})^\alpha},
\end{equation}
and (a) follows from the independence of ${ \frac{|h_{B_*E_e}|^2}{|h_{UE_e}|^2}; E_e\in\Phi}$, (b) holds since the CDF
\begin{equation}
F_\nu(\nu) = \mathbb{P}\left(\frac{|h_{B_*E_e}|^2}{|h_{UE_e}|^2}<\nu\right) = \frac{\nu}{\nu+d^\alpha_{UE}/d^\alpha_{BE}},
\end{equation}
and (c) holds for the probability generating functional lemma \cite{M.H12}. Then by using \eqref{14} and \eqref{16}, the secrecy outage probability of the FD UE can be written as
\begin{equation}\label{mm}
\begin{split}
P^{(F)}_{so} &\leq 1- \int_0^\infty f_{X}(x)F_{Y}\left(\frac{x}{\beta}\right)\rmd x,
\end{split}
\end{equation}
which has been shown in Proposition 3.

\section*{Appendix IV}
According to the definition of secrecy outage probability \eqref{4}, \eqref{5} and \eqref{6} with $\varpi = 0$, we can obtain the secrecy outage probability as followed
\begin{equation}\label{23}
\begin{split}
P^{(H)}_{so} &= \mathbb{P}\left(\frac{\underset{k\in (1...K)}{\max}\left(\frac{|h_{B_kU}|^2}{d_{BU}^\alpha}\right)}{\underset{e\in \Phi}\sum\left(\frac{|h_{B_*E_e}|^2}{d_{BE_e}^\alpha}\right)}<\beta\right)\\
& = \mathbb{P}\left(\underset{k\in (1...K)}{\max}\left(\frac{|h_{B_kU}|^2}{d_{BU}^\alpha}\right)<\beta\sum_{e\in\Phi}\left(\frac{|h_{B_*E_e}|^2}{d_{BE_e}^\alpha}\right)\right)\\
& = \sum^K_{k=0}C^k_K(-1)^k\int_0^\infty e^{-k\beta zd^\alpha_{BU}}f_Z(z)\rmd z\\
& = \sum^K_{k=0}C^k_K(-1)^k\mathbb{E}\left[e^{-sZ}\right]|_{s=k\beta d^\alpha_{BU}}
\end{split}
\end{equation}
where $Z =\underset{e\in \Phi}\sum\left(\frac{|h_{B_*E_e}|^2}{d_{BE_e}^\alpha}\right)$ and $\mathbb{E}\left[e^{-sZ}\right]|_{s=k\beta d^\alpha_{BU}}$ is given by
\begin{equation}\label{24}
\begin{split}
\mathbb{E}&\left[e^{-sZ}\right]|_{s=k\beta d^\alpha_{BU}}=\mathbb{E}\left[\underset{e\in \Phi}\prod e^{-k\beta d^\alpha_{BU}|h_{B_*E_e}|^2d_{BE_e}^{-\alpha}}\right]\\
& = \mathbb{E}_\Phi\left[\underset{e\in \Phi}\prod \mathbb{E}_{|h_{B_*E_e}|^2}\left[e^{-k\beta d^\alpha_{BU}|h_{B_*E_e}|^2d_{BE_e}^{-\alpha}}\right]\right]\\
& \overset{(a)}{=} \mathbb{E}_\Phi\left[\underset{e\in \Phi}\prod \int_0^\infty e^{-k\beta d^\alpha_{BU}td_{BE_e}^{-\alpha}}e^{-t}\rmd t\right]\\
& = \mathbb{E}_\Phi\left[\underset{e\in \Phi}\prod \frac{1}{1+k\beta(d_{BU}/d_{BE_e})^\alpha}\right]\\
& \overset{(b)}{=} \textrm{exp}\left(-\rho_E\int_0^{2\pi}\int_0^{R}\left(1-\frac{1}{1+k\beta(d_{BU}/r)^\alpha}\right)r\rmd r\rmd \theta\right)\\
& = \exp\!\left(-\pi R^2\rho_E F\!\left(1,\frac{2}{\alpha};1+\frac{2}{\alpha};-\frac{R^\alpha}{k\beta d_{BU}^\alpha}\right)\right),
\end{split}
\end{equation}
where, for brevity and ease of exposition, we let $t = |h_{B_*E_e}|^2$ in (a) and the PDF of $t$ is $e^{-t}$, $F(a,b;c;z)$ denotes the Gaussian hypergeometric function, and (b) holds for the probability generating functional lemma \cite{M.H12}.

\section*{Appendix V}
According to the definition of secrecy outage probability in \eqref{4}, \eqref{5} and \eqref{6} with $\varpi = 1$, modeling the residual self-interference as AWGN noise \cite{B.D14,M.J11} and ignoring the noise at ED as in \cite{X.Z13,C.W15,X.Z131}, we can obtain the secrecy outage probability as follows

\begin{equation}\label{26}
\begin{split}
P^{(F)}_{so} &\leq \mathbb{P}\left(\frac{\frac{P_B}{2}\underset{k\in (1...K)}{\max}\left(\frac{|h_{B_kU}|^2}{d^\alpha_{BU}}\right)}{\underset{e\in \Phi}{\sum}\left(\frac{P_B\frac{|h_{B_*E_e}|^2}{d^\alpha_{BE_e}}}{P_U\frac{|h_{UE_e}|^2}{d^\alpha_{UE_e}}}\right)}<\beta\right)\\
& = \mathbb{P}\left(\underset{k\in (1...K)}{\max}\left(\frac{|h_{B_kU}|^2}{d_{BU}^\alpha}\right)<\frac{2\beta}{P_U}\underset{e\in \Phi}{\sum}\left(\frac{\frac{|h_{B_*E_e}|^2}{d^\alpha_{BE_e}}}{\frac{|h_{UE_e}|^2}{d^\alpha_{UE_e}}}\right)\right)\\
& = 1+\sum^K_{k=1}C^k_K(-1)^k\int_0^\infty e^{-\frac{2k\beta}{P_U} zd^\alpha_{BU}}f_Z(z)dz\\
& = 1+\sum^K_{k=1}C^k_K(-1)^k\mathbb{E}\left[e^{-sZ}\right]|_{s=\frac{2k\beta}{P_U} d^\alpha_{BU}},
\end{split}
\end{equation}
where $Z = \underset{e\in \Phi}{\sum}\left(\frac{\frac{|h_{B_*E_e}|^2}{d^\alpha_{BE_e}}}{\frac{|h_{UE_e}|^2}{d^\alpha_{UE_e}}}\right)$ and $\mathbb{E}\left[e^{-sZ}\right]|_{s=\frac{2k\beta}{P_U} d^\alpha_{BU}}$ can be obtained as \eqref{27} at the top of the next page.
\begin{figure*}
\begin{equation}\label{27}
\begin{split}
\mathbb{E}&\left[e^{-sZ}\right]|_{s=\frac{2k\beta}{P_U} d^\alpha_{BU}}=\mathbb{E}\left[\underset{e\in \Phi}\prod e^{-\frac{2k\beta}{P_U} d^\alpha_{BU}\frac{td_{BE_e}^{-\alpha}}{\left(\sqrt{d_{BE_e}^2+d_{BU}^2-2d_{BE_e}d_{BU}\textrm{cos}(\theta)}\right)^{-\alpha}}}\right]\\&
= \mathbb{E}_\Phi\left[\underset{e\in \Phi}\prod \mathbb{E}_{t}\left[e^{-\frac{2k\beta}{P_U} d^\alpha_{BU}\frac{td_{BE_e}^{-\alpha}}{\left(\sqrt{d_{BE_e}^2+d_{BU}^2-2d_{BE_e}d_{BU}\textrm{cos}(\theta)}\right)^{-\alpha}}}\right]\right]\\
& \overset{(a)}{=} \mathbb{E}_\Phi\left[\underset{e\in \Phi}\prod \int_0^\infty e^{-\frac{2k\beta}{P_U} d^\alpha_{BU}t\left(\frac{d_{BE_e}}{\sqrt{d_{BE_e}^2+d_{BU}^2-2d_{BE_e}d_{BU}\textrm{cos}(\theta)}}\right)^{-\alpha}}\frac{1}{(1+t)^2}\rmd t\right]
\overset{(b)}{=} \textrm{exp}\left(-\rho_E\int_0^{R}\int_0^{2\pi}Ae^{A}\mathsf{E}_1\left(A\right)r\rmd \theta \rmd r\right)\\ & \overset{(c)}{\simeq} \textrm{exp}\left(-\rho_E\left(\int_0^{\varrho}\int_0^{2\pi}(1-1/A)r\rmd r\rmd\theta+\int_\varrho^{R}\int_0^{2\pi}A(A+1)\left(A-\ln (A)-\kappa\right)r\rmd r\rmd\theta\right)\right)\\ &\overset{(d)}{=} \textrm{exp}\left(-\rho_E\left(\pi\varrho^2-\frac{\pi P_U}{2k\beta}\left(\ln \left(\frac{1}{1-\left(\frac{\varrho}{d_{BU}}\right)^2}\right)+\left(\frac{\varrho}{d_{BU}}\right)^2\right)+\Omega(\beta;d_{BU},R,A_0) \right) \right)_.
\end{split}
\end{equation}
\hrule
\end{figure*}
For brevity and ease of exposition, we let $t = \frac{|h_{B_*E_e}|^2}{|h_{UE_e}|^2}$ in (a) and the PDF of $t$ is $1/(1+t)^2$, $A=\frac{2k\beta}{P_U} d^\alpha_{BU}\left(\frac{r}{\sqrt{r^2+d_{BU}^2-2rd_{BU}\textrm{cos}(\theta)}}\right)^{-\alpha}$, $A_0 = \frac{2k\beta}{P_U} d^2_{BU}$, $\Omega(\beta;d_{BU},R,A_0)$ is given as \eqref{omega} and (b) holds for the probability generating functional lemma \cite{M.H12}. In (c), the first double integral can be approximately obtained by using asymptotic (divergent) series \cite{N.B86} and the second double integral can be approximated by using the Taylor series \cite{C.M78}, and (d) holds when $\alpha = 2$.

\section*{Acknowledgements}
The authors wish to thank Prof. C. Dettmann, Dr. K. Koufos and Dr. D. Simmons for their input. We also would like to thank the anonymous reviewers and the editor for their constructive comments.


\bibliographystyle{ieeetr}

\begin{thebibliography}{9}
\bibitem{A.D75}
A. D. Wyner, ``The wire-tap channel", \textit{Bell Syst. Tech. J.}, vol. 54, pp. 1355-1387, Jan. 1975.

\bibitem{E.T08}
E. Tekin and A. Yener, ``The general Gaussian multiple access and two-way wire-tap channels: achievable rates and cooperative jamming", \textit{IEEE Trans. Inform. Theory}, vol. 54, pp. 2735-2751, June 2008.

\bibitem{P.P09}
P. Popovski and O. Simeone, ``Wireless Secrecy in Cellular Systems With Infrastructure-aided Cooperation", \textit{IEEE Trans. Inform. Forensics and Security}, vol. 4, no. 4, pp. 242-256, June 2009.

\bibitem{G.C14}
G. Chen, Z. Tian, Y. Gong, Z. Chen and J. A. Chambers, ``Max-ratio relay selection in secure buffer-aided cooperative wireless networks", \textit{IEEE Trans. Inform. Forensics and Security}, vol. 9, no. 4, pp. 719-729, Apr. 2014.

\bibitem{T.L09}
T. Liu and S. Shamai, ``Max-ratio relay selection in secure buffer-aided cooperative wireless networks", \textit{IEEE Trans. Information Theory}, vol. 55, no. 6, pp. 2547-2553, June 2009.

\bibitem{H.W09}
G. Chen, Z. Tian, Y. Gong, Z. Chen and J. A. Chambers, ``The Capacity Region of the Degraded Multiple-Input Multiple-Output Compound Broadcast Channel", \textit{IEEE Trans. Inf. Theory}, vol. 55, no. 11, pp. 5011-5023, Nov. 2009.


\bibitem{G.C15}
G. Chen, Y. Gong, P. Xiao and J. A. Chambers, ``Physical Layer Network Security in the Full-Duplex Relay System", \textit{IEEE Trans. Inform. Forensics and Security}, vol. 10, no. 3, pp. 574-583, Apr. 2015.

\bibitem{G.C16}
G. Chen, Y. Gong, P. Xiao and J. A. Chambers, ``Dual Antenna Selection in Secure Cognitive Radio Networks", \textit{IEEE Trans. Veh. Technol.}, vol. 65, no. 10, pp. 7993-8002, Apr. 2015.


\bibitem{L.D09}
L. Dong, Z. Han, A. P. Petropulu and H. V. Poor, ``Amplify-and-forward based cooperation for secure wireless communications", \textit{in Proc. IEEE Int. Conf. Acoustics, Speech, and Signal Processing, Taipei, Taiwan}, Apr. 2009.

\bibitem{M.H08}
M. Haenggi, ``The secrecy graph and some of its properties", \textit{in Proc. IEEE Int. Symp. Inf. Theory, Toronto, Canada}, July 2008.


\bibitem{P.C08}
P. C. Pinto, J. Barros and M. Z. Win, ``Physical-layer security in stochastic wireless networks", \textit{in Proc. IEEE Int. Conf. Commun. Syst.,Guangzhou, China,}, Nov. 2008.


\bibitem{X.Z11}
X. Zhou, R. K. Ganti, J. G. Andrews and A. Hjorungnes, ``On the Throughput Cost of Physical Layer Security in Decentralized Wireless Networks", \textit{IEEE Trans. on Wireless Commun.}, vol. 10, no. 8, pp. 2764-2775, Aug. 2011.

\bibitem{G.G14}
G. Geraci, S. Singh, J. G. Andrews and J. Yuan and I. B. Collings, ``Secrecy Rates in Broadcast Channels with Confidential Messages and External Eavesdroppers", \textit{IEEE Trans. on Wireless Commun.}, vol. 13, no. 5, pp. 2931-2943, May 2014.

\bibitem{T.X14}
T. X. Zheng,H. M. Wang and Q. Yin, ``On Transmission Secrecy Outage of a Multi-Antenna System With Randomly Located Eavesdroppers", \textit{IEEE Commun. Lett.}, vol. 18, no. 8, pp. 1299-1302, Aug. 2014.

\bibitem{M.G11}
M. Ghogho and A. Swami, ``Physical-Layer Secrecy of \textsc{MIMO} Communications in the Presence of a Poisson Random Field of Eavesdroppers", \textit{in Proc. IEEE ICC}, June 2011.

\bibitem{T.X15}
T. X. Zheng, H. M. Wang, J. Yuan, D. Towsley and M. H. Lee, ``Multi-Antenna Transmission With Artificial Noise Against Randomly Distributed Eavesdroppers", \textit{IEEE Trans. on Commun.}, vol. 63, no. 11, pp. 4347-4362, Nov. 2015.


\bibitem{G.C13}
G. Chen, Y. Gong and J. A. Chambers, ``Study of relay selection in a multi-cell cognitive network", \textit{IEEE Wireless Commun. Lett.}, vol. 2, no. 4, pp. 435-438, Aug. 2013.


\bibitem{P.C122}
P. C. Pinto, J. Barros and M. Z. Win, ``Secure Communication in Stochastic Wireless Networks-Part \textsc{II}: Maximum Rate and Collusion", \textit{IEEE Transactions on Information Forensics and Security}, vol. 7, no. 4, pp. 139-147, Feb. 2012.


\bibitem{X.Z111}
X. Zhou, R. K. Ganti and J. G. Andrews, ``Secure Wireless Network Connectivity with Multi-Antenna Transmission", \textit{IEEE Transactions on Wireless Communications}, vol. 10, no. 2, pp. 425-430, Feb. 2011.


\bibitem{P.C121}
P. C. Pinto, J. Barros and M. Z. Win, ``Secure Communication in Stochastic Wireless Networks-Part \textsc{I}: Connectivity", \textit{IEEE Transactions on Information Forensics and Security}, vol. 7, no. 1, pp. 125-138, Feb. 2012.

\bibitem{X.Z13}
X. Zhang, X. Zhou and M. R. McKay, ``Enhancing Secrecy With Multi-Antenna Transmission in Wireless Ad Hoc Networks", \textit{IEEE Transactions on Information Forensics and Security}, vol. 8, no. 11, pp. 1802-1814, Nov. 2013.


\bibitem{C.W15}
C. Wang, H. M. Wang, X. G. Xia and C. Liu, ``Uncoordinated Jammer Selection for Securing \textsc{SIMOME} Wiretap Channels: A Stochastic Geometry Approach", \textit{IEEE Trans. on Wireless Commun.}, vol. 14, no. 5, pp. 2596-2612, May 2015.

\bibitem{X.Z131}
X. Zhang, X. Zhou and M. R. McKay, ``On the Design of Artificial-Noise-Aided Secure Multi-Antenna Transmission in Slow Fading Channels", \textit{IEEE Trans. on Veh. Tech.}, vol. 62, no. 5, pp. 2170-2181, June 2013.

\bibitem{S.H14}
S. Hong, J. Brand, J. Choi, M. Jain, J. Mehlman, S. Katti and P. Levis, ``Applications of self-interference cancellation in 5\textsc{G} and beyond", \textit{IEEE Commun. Magazine}, vol. 52, no. 2, pp. 114-121, Feb. 2014.


\bibitem{M.H12}
M. Haenggi, ``Stochastic Geometry for Wireless Networks", \textit{Cambridge Univ. Press}, 2012.


\bibitem{I.S07}
I. S. Gradshteyn and I. M. Ryzhik, ``Table of Integrals, Series, and Products", \textit{Elsevier Academic Press}, 7th ed. 2007.


\bibitem{M.A72}
M. Abramowitz and I. A. Stegun, ``Handbook of Mathematical Functions with Formulas, Graphs, and Mathematical Tables", \textit{Dover, New York, 9th ed.}, 7th ed. 1972.

\bibitem{B.D14}
B. Debaillie, D. J. Broek, C. Lavin, B. Liempd, E. A. M. Klumperink, C. Palacios, J. Craninckx, B. Nauta and A. Parssinen, ``Analog/\textsc{RF} Solutions Enabling Compact Full-Duplex Radios", \textit{IEEE J. Sel. Areas Commun.}, vol. 32, no. 9, pp. 1662-1673, Oct. 2014.


\bibitem{M.J11}
M. Jain, J. I. Choi, T. Kim, D. Bharadia, S. Seth, K. Srinivasan, P. Levis, S. Katti and P. Sinha, ``Practical, real-time, full duplex wireless", \textit{in Proc. ACM. MobiCom.}, Sep. 2011.


\bibitem{N.B86}
N. Bleistein and R. A. Handelsman, ``Asymptotic Expansions of Integrals", \textit{Dover Publications., New york}, 1986.


\bibitem{C.M78}
C. M. Bender and S. A. Orszag, ``Advanced mathematical methods for scientists and engineers", \textit{McGraw–Hill}, 1978.

\end{thebibliography}

\end{document}